\documentclass[conference]{IEEEtran} 

\usepackage{setspace} 

\usepackage{stfloats}
\usepackage{graphicx}
\usepackage{amsmath}
\usepackage{amssymb}

 \newcommand{\be}{\begin{equation}}
 \newcommand{\ee}{\end{equation}}
 \newcommand{\ba}{\begin{array}}
\newcommand{\ea}{\end{array}}

\renewcommand{\thesubsubsection}{\thesubsection.\arabic{subsubsection}}

\renewcommand{\baselinestretch}{1.0}
\usepackage{caption2}

\title{Iterative Hybrid Precoder and Combiner Design for mmWave MIMO-OFDM Systems
\thanks{This paper is supported in part by the Natural Science Foundation of Liaoning Province (Grant No. 2015020043) and in part Fundamental Research Funds for the Central Universities (Grant No. DUT 15RC(3)121).}}

\author{ \IEEEauthorblockN{Wenfei Liu$^{\dag}$, Ming Li$^{\dag}$,  Xiaowen Tian$^{\dag}$, Zihuan Wang$^{\dag}$, and Qian Liu$^{ \ddag}$
\vspace{-0.0 cm} }\\
\IEEEauthorblockA{$^{\dag}$School of Information and Communication Engineering   \\  Dalian University of Technology, Dalian, Liaoning 116024, China  \\ E-mail: \texttt{\{liuwenfei,tianxw,wangzihuan\}@mail.dlut.edu.cn, mli@dlut.edu.cn}}

\IEEEauthorblockA{$^{\ddag}$   School of Computer Science and Technology \\  Dalian University of Technology, Dalian, Liaoning 116024, China \\ E-mail: \texttt{qianliu@dlut.edu.cn}} }

\begin{document}

\pagestyle{empty}

 \maketitle

\vspace{-0.0 cm}

\begin{abstract}
This paper investigates the problem of hybrid precoder and combiner design for multiple-input multiple-output (MIMO) orthogonal frequency division multiplexing (OFDM) systems operating in millimeter-wave (mmWave) bands.
We propose a novel iterative scheme to
design the codebook-based analog precoder and combiner in forward and reverse channels.
During each iteration, we apply compressive sensing (CS) technology to efficiently estimate the equivalent MIMO-OFDM mmWave channel.
Then, the analog precoder or combiner is obtained
based on the orthogonal matching pursuit (OMP) algorithm to alleviate the interference between different data streams
as well as maximize the spectral efficiency.
The digital precoder and combiner are finally obtained based on the effective baseband channel
to further enhance the spectral efficiency.
Simulation results demonstrate the proposed iterative hybrid precoder and combiner algorithm has
significant performance advantages.
\end{abstract}

\vspace{-0.0 cm}

\section{Introduction}
Millimeter wave (mmWave) communications can provide high data rates by leveraging
the large unexploited bandwidths ranged from 30GHz to 300GHz,
which makes mmWave communication a promising  candidate to solve the spectrum congestion problem in
the future wireless communication networks \cite{Fre}-\cite{5G}.
However, compared with the conventional frequency bands, the propagation loss in
the mmWave band is much more severe  due to rain attenuation and low penetration.
Thanks to the small wavelength of mmWave signals which enables a large array to be
packed into a small physical dimension, mmWave communications with massive MIMO systems can
provide the significant beamforming gains to overcome
severe path loss of mmWave channel as well as enable the transmission of multiple data streams \cite{Alt}.

In the conventional MIMO systems, full-digital precoders and combiners   accomplished
in digital-domain can adjust both magnitude and phase of the transmit and receive signals.
However, these full-digital precoders and combiners require
a large number of expensive and energy-intensive radio frequency
(RF) chains, analog-to-digital converters (ADCs), and digital-to-analog
converters (DACs)  which make full-digital precoding
and combining schemes impractical in mmWave communication systems \cite{RF}.
Recently, hybrid architectures have been considered as an emerging technique
to solve this issue.
The hybrid beamformer can achieve high spectral efficiency and maintain low cost and power assumption
compared with the traditional MIMO systems \cite{hybrid}.
The hybrid precoding/combining architectures can
apply high-dimensional RF precoder with large number of analog phase shifters
to compensate the large path loss at mmWave bands.
Moreover,  a small number of RF chains for low-dimensional digital precoder can provide
necessary flexibility to perform spatial multiplexing.
The hybrid precoder design problem is usually formulated
 to solve various matrix factorization problems with constant modulus
constraints of the analog precoder, which is imposed by the phase shifters.
Particularly,  according to the special characteristic of mmWave channel,
a codebook-based hybrid precoder design technique has been widely used,
where the columns of the analog precoder
are selected from pre-specified vectors, such as array response
vectors of the channel and discrete Fourier transform (DFT) beamformers.

Most prior works have been devoted to investigating hybrid precoding and combing algorithms in narrowband channels \cite{spatially}-\cite{narrow2}.
In \cite{spatially},\cite{narrow1},  the  spatial structure of mmWave channels is exploited
to formulate transmit precoding and receive combining problems as an OMP algorithm.
In \cite{narrow2}, authors propose an iterative algorithm
which updates the phases of the phase shifters of the RF precoder and combiner.
Extensions to  wideband mmWave hybrid precoding systems have been investigated in \cite{wideband1}-\cite{wideband3}.
\cite{wideband1} demonstrates  the feasibility for  millimeter-wave mobile broadband (MMB)
to achieve gigabit-per-second data rates at a distance up to 1 km in an urban mobile environment.
In \cite{wideband2}, a multi-beam transmission diversity scheme for single stream transmission in MIMO-OFDM system is proposed.
In \cite{wideband3}, the authors consider a limited feedback hybrid precoding system to design precoders and combiners.



In this paper, we consider a wideband mmWave MIMO-OFDM system
with unknown channel state information (CSI).
We propose a novel iterative hybrid precoder and combiner design in both forward and reverse channel.
A CS-based channel estimation is firstly  utilized  to estimate the effective channel.
Then, based on the effective channel, the analog precoder or combiner
is obtained using OMP algorithm. Finally,
the digital  precoder and combiner
are  obtained to further suppress the interference and maximize the spectral efficiency.
Simulation results show that our proposed algorithm can achieve
significant performance improvement.

The following notation is used throughout this paper.
$(\cdot)^T$, $(\cdot)^H$, and $(\cdot)^*$ are the transpose,
conjugate transpose, and  conjugate of a matrix, respectively.
$\mathbb{E}\{\cdot \}$ denotes statistical expectation.
$\mathbb{C}^{M \times N}$ is the set of  $M \times N$ matrices with complex entries.
$\mathbf{I}_N$ represents an $N \times N$ identity matrix.
$|\cdot|$ , $\| \cdot \|$, and $\|\cdot\|_F$
are the scalar magnitude, vector norm,
and Frobenius norm, respectively.
$[\cdot]_{r, :}$ and $ [ \cdot]_{:,c}$ are the $r$th row and $c$th column of a matrix.


\begin{figure*}[!tp]
   \begin{center}
    \vspace{-0.3cm}
        \includegraphics[width= 5.7 in]{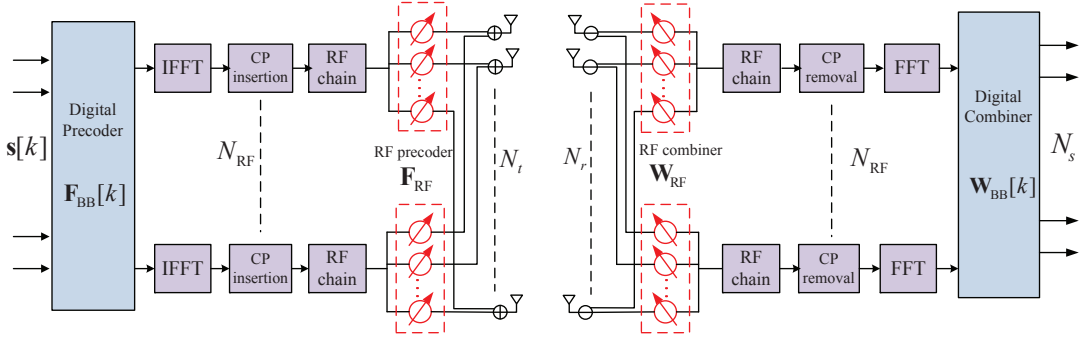}
         \vspace{-0.2 cm}
         \captionsetup{justification=centering}
    \caption{The hybrid precoding and combining architecture in a  MIMO-OFDM system.}
   \label{system1}
   \end{center}
\end{figure*}

\section{System and Problem Formulation}

In this section, we present the system and channel models for  mmWave MIMO-OFDM communications with hybrid
precoder and combiner architecture.

\subsection{System Model}
We consider a MIMO-OFDM system with $N$ subcarriers as shown in Fig. \ref{system1}.
A transmitter with $N_t$ transmit antennas and $N_{\mathrm{RF}}$ RF chains transmits  $N_s$ data streams to
a receiver which has $N_r$ receive antennas and $N_{\mathrm{RF}}$ RF chains.
We assume the number of RF chains is subject to the constraints
 $ N_\mathrm{RF}\leq N_t$ and $ N_{\mathrm{RF}}\leq N_r$.
 The number of data streams is constrained as $N_s= N_{\mathrm{RF}}$.

At the transmitter, let $\mathbf{s}[k]$,  $k=1,\ldots,N$,
be  the $N_s \times 1$ transmitted vector at the $k$th subcarrier,
$\mathbb{E}\{ \mathbf{s}[k]\mathbf{s}^H[k]\}=\frac{1}{N_s}\mathbf{I}_{N_s}$.
 The data stream is firstly
precoded by the $N_{\mathrm{RF}}\times N_s$ digital precoding matrix $\mathbf{F}_{\mathrm{BB}}[k]$,
and then transformed to the time-domain by inverse fast Fourier transform (IFFT) operation.
After cyclic prefix (CP) insertion, the transmitted signal is precodered by the $N_t \times N_\mathrm{RF}$
analogy precoder $\mathbf{F}_\mathrm{RF}$.
Note that  the digital precoding matrix $\mathbf{F}_{\mathrm{BB}}[k]$ can be different for each subcarrier,
the analog precoding matrix $\mathbf{F}_\mathrm{RF}\in \mathcal{F}$ is the same for all subcarriers
due to the special hybrid precoding constructure, where  $\mathcal{F}$ is the codebook for
 the analog precoders which are implemented
by analog components like phase shifters, i.e. a set of $N_t \times 1$ vectors with
 quantized phases and constant magnitude entries.
The transmit signal at the $k$th subcarrier can be expressed as
\be  \mathbf{x}[k]= \sqrt{P} \mathbf{F}_\mathrm{RF} \mathbf{F}_{\mathrm{BB}}[k] \mathbf{s}[k], k=1,\ldots,N,  \ee
where $P$ represents the average transmit power.
The main difference between the OFDM-based hybrid precoding and conventional fully-digital precoding
is that $\mathbf{F}_\mathrm{RF}$ is applied in the time-domain and the same for all subcarriers,
while the baseband precoder $\mathbf{F}_{\mathrm{BB}}[k]$ is performed on each subcarrier in the frequency-domain.


At the receiver, the received signal is combined by the $N_r \times N_{\mathrm{RF}}$ RF combining matrix $\mathbf{W}_\mathrm{RF}$.
The constraint of RF combiner $\mathbf{W}_\mathrm{RF}$ is similar to the RF precoder $\mathbf{F}_\mathrm{RF}$,
i.e. $\mathbf{W}_\mathrm{RF} \in \mathcal{W}$, where $ \mathcal{W}$ is the set of feasible RF combiners.
After RF combining, the CP is removed, and then
the time-domain signal is transformed to frequency-domain by the FFT operation.
Finally, the $N_{\mathrm{RF}} \times N_s$  digital combing matrix
$\mathbf{W}_{\mathrm{BB}}[k]$ is employed to  process the signal at the $k$th  subcarrier. Let $\mathbf{H}[k]\in \mathbb{C}^{N_r \times N_t}$ denote the
channel at the $k$th subcarrier, the received signal after combing at the $k$th subcarrier can be expressed as
\be
\begin{split}
\mathbf{y}[k]= & \sqrt{P}\mathbf{W}^H_{\mathrm{BB}}[k] \mathbf{W}^H_\mathrm{RF} \mathbf{H}[k] \mathbf{F}_\mathrm{RF} \mathbf{F}_{\mathrm{BB}}[k] \mathbf{s}[k] \\
&+ \mathbf{W}^H_{\mathrm{BB}}[k] \mathbf{W}^H_\mathrm{RF} \mathbf{n}[k], k=1, \ldots, N,
\label{y}
\end{split}
\ee
where $\mathbf{n}[k]\sim \mathcal{CN}(0,\sigma_n^2\mathbf{I})$ is the Gaussian noise vector at the $k$th subcarrier.

\subsection{Channel Model}

We consider a geometric channel model for wideband mmWave channel. 
For the ease of description, we will use linear antenna array in the channel model.
In the $N$-subcarrier OFDM system, the delay-domain channel with $L$ number of paths can be expressed as
\be
\mathbf{H}[d]=\sum_{l=1}^{L} \alpha_l f(d T_s-\tau_l)\mathbf{a}_r(\varphi_l) \mathbf{a}_t^H(\phi_l),
\label{Hd}
\ee
where $\alpha_l$ is the complex path gain of the $l$th channel path,
$f(\tau)$ denotes the raised cosine pulse filter at $\tau$, $T_s$ is sampling period,
$\tau_l \in \mathbb{R}$ is the delay of the $l$th path.
$\varphi_l$ and $\phi_l$ are the angles of arrival (AoA) and departure (AoD), respectively.
$\mathbf{a}_r(\varphi_l)$ and $\mathbf{a}_t(\phi_l)$ denote the antenna array response of the
transmitter and receiver, respectively,
\be \mathbf{a}_r=[1, e^{j\pi \mathrm{sin}(\varphi_l)},\ldots, e^{j\pi (N_r-1) \mathrm{sin}(\varphi_l)}]^T, \label{ar} \ee
\be \mathbf{a}_t=[1, e^{j\pi \mathrm{sin}(\phi_l)},\ldots, e^{j\pi (N_t-1) \mathrm{sin}(\phi_l)}]^T .  \label{at} \ee

After $N$-point FFT of the delay-domain channel $\mathbf{H}[d]$,
we can obtain the frequency-domain channel at the $k$th  subcarrier
\be
\begin{split}
\mathbf{H}[k]=& \sum_{d=0}^{N-1} \mathbf{H}[d] e^{-j \frac{2\pi k}{N}d} \\
=&\sum_{d=0}^{N-1}\sum_{l=1}^{L} \alpha_l f(d T_s-\tau_l)\mathbf{a}_r(\varphi_l) \mathbf{a}_t^H(\phi_l) e^{-j \frac{2\pi k}{N}d },
\end{split}
\label{Hf}
\ee
where $k=0,1,\ldots, N-1,$ denotes the $k$th subcarrier in the OFDM systems.

\subsection{Problem Formulation}

In this paper, we consider the problem of hybrid precoder and combiner
design in mmWave MIMO-OFDM systems.
We first present the CS-based channel estimation
to obtain the effective channel information.
Then, with the aid of the CS-based channel estimation, we propose an iterative  hybrid precoder and combiner design scheme
aiming at maximizing  the spectral efficiency
which can be expressed as
\begin{small}
\be
\begin{split}
& \mathcal{I} = \frac{1}{N}\sum_{k=1}^{N} \mathrm{log}_2 \Bigg( \bigg|\mathbf{I}_{N_s} + \frac{P}{N_s}\mathbf{R}_n^{-1} \mathbf{W}_\mathrm{{BB}}^H[k]
\mathbf{W}^H_\mathrm{{RF}}\mathbf{H} [k]\mathbf{F}_\mathrm{{RF}} \\
& \times \mathbf{F}_\mathrm{{BB}}[k] \mathbf{F}_\mathrm{{BB}}^H[k] \mathbf{F}_\mathrm{RF}^H \mathbf{H}^H[k] \mathbf{W}_\mathrm{{RF}}\mathbf{W}_\mathrm{{BB}}[k] \bigg| \Bigg),
 k=1,\ldots, N.
\label{sumrate}
\end{split}
\ee
\end{small}
where  $\mathbf{R}_n \triangleq \sigma_n^2 \mathbf{W}_\mathrm{{BB}}^H[k] \mathbf{W}^H_\mathrm{{RF}}  \mathbf{W}_\mathrm{{RF}}\mathbf{W}_\mathrm{{BB}}[k]$.

The optimization problem of (\ref{sumrate}) is obviously a non-convex problem.
Note that the forward channel (transmitter-to-receiver) and the reverse
channel (receiver-to-transmitter) are identical in a reciprocal time division duplex (TDD) system.
 Motivated by this fact, we propose a  forward-reverse iterative CS-based hybrid beamformer design algorithm.
 At each iteration,
the transmitter and receiver conditionally determine their optimal beam vectors based on the estimated
forward or reverse effective channel information obtained by CS technique. The detailed algorithm is presented in the next section.

\section{Hybrid Precoder and Combiner Design}

\subsection{Analog Combiner Design with CS-based Forward Channel Estimation}

%
%
%

%
%

The transmitter firstly transmits some training symbols in order to let the receiver estimate the forward channel.
We assume different RF chains are independent with each other and  the training signals are transmitted one RF chain by another.
The received signal at the $k$th subcarrier for the $i$th RF chain is
\be
\mathbf{Y}_r^i[k]=  \mathbf{H}_r[k] \mathbf{f}_{\mathrm{RF}_i} \mathbf{s},  i=1, \ldots, N_{\mathrm{RF}},
\label{yr}
\ee
where $\mathbf{Y}_r^i[k] \in \mathbb{C}^{N_r \times 1}$ represents the received signal  for the $i$th RF chain,
$\mathbf{s}=[1, 1, \dots, 1]^T$ represents the training signal for all RF chains.
  The subscript $``r"$ represents the receiver.
 In the initial iteration, an random  beam vector  $\mathbf{f}_{\mathrm{RF}_i}$ is adopted in each RF chains
 since we do not know the exact position of the receiver.
 With the analog precoder vector $\mathbf{f}_{\mathrm{RF}_i}$, the frequency-domain channel vector
 in the first RF chain can be expressed as
 \be
 \begin{split}
\mathbf{h}_r^1[k] = & \mathbf{H}_r[k] \mathbf{f}_{\mathrm{RF}_1},\\
=&\sum_{d=0}^{N-1}\sum_{l=1}^{L} \alpha_l f(d T_s-\tau_l)e^{-j \frac{2\pi k}{N}d }  \mathbf{a}_r(\varphi_l) \mathbf{a}_t^H(\phi_l) \mathbf{f}_{\mathrm{RF}_1},\\
=& \sum_{d=0}^{N-1}\sum_{l=1}^{L} \underbrace{\alpha_l  \beta_1 f(d T_s-\tau_l)e^{-j \frac{2\pi k}{N}d } }_{\gamma_{d,l,k}} \mathbf{a}_r(\varphi_l), k=1, \ldots, N,
\label{hr1}
\end{split}
 \ee
  where $\beta_i \triangleq \mathbf{a}_t^H(\phi_l) \mathbf{f}_{\mathrm{RF}_i}$, $i=1, \ldots, N_\mathrm{RF}$.

We define a dictionary $\mathbf{\Psi}_r$ shown in (\ref{PHIr}) at the top of the next page,
where $\theta \in [-90^\circ, 90^\circ]$ is the angle resolution,
$\mathbf{\Psi}_r$ has $L_r=180/\theta$ columns.
The dictionary $\mathbf{\Psi}_r$ can cover the whole angle range
since $\theta \in [-90^\circ, 90^\circ]$ and produce all possible
values of $\mathrm{sin}(\theta )$.
\begin{figure*}[ht]
\begin{small}
\be \mathbf{\Psi}_r   =  \left[ \begin{array}{c c c c c c c}
 1 & \ldots & 1 & 1 & 1 & \ldots & 1\\
e^{j \pi sin(-90^\circ+\theta)} & \ldots & e^{j \pi sin(-\theta) }& 1 & e^{j \pi sin(\theta)} & \ldots & e^{j \pi sin(-90^\circ)}\\
e^{j 2\pi sin(-90^\circ+\theta)} & \ldots & e^{j 2\pi sin(-\theta) }& 1 & e^{j 2\pi sin(\theta)} & \ldots & e^{j 2\pi sin(-90^\circ)}\\
\vdots & \ldots &\vdots &\vdots &\vdots &\ldots &\vdots \\
e^{j (N_r-1) \pi sin(-90^\circ+\theta)} & \ldots & e^{j(N_r-1) \pi sin(-\theta) }& 1 & e^{j (N_r-1)\pi sin(\theta)} & \ldots & e^{j (N_r-1)\pi sin(-90^\circ)}
 \end{array}
\right].
\label{PHIr} \vspace*{-0.0 cm} \ee
\end{small}
\end{figure*}
Then, we can represent the effective channel $\mathbf{h}_r^1[k] $ as
\be
\mathbf{h}_r^1[k] =\mathbf{\Psi}_r \mathbf{x}_r^1[k], k=1, 2, \ldots, N.
\label{HX}
\ee
If the $i$th column of $\mathbf{\Psi}_r $ is equal to  $\mathbf{a}_r(\varphi_l)$,
then the $i$th entry of $\mathbf{x}_r^1[k]$ is set to $\gamma_{d,l,k}$.
This means that, the expansion vector  $\mathbf{x}_r^1[k]$ of $\mathbf{h}_r^1[k]$
is sparse, which enables us utilize CS to estimate the effective channel.

After the training for the first RF chain, the other training procedures are
similar as above. Let $\widetilde{\mathbf{H}}_r[k]$  be the  effective channel matrix,
$\mathbf{X}_r[k]\in \mathbb{C}^{L_r \times N_{\mathrm{RF}}}$ be the sparse matrix,
and each column of $\mathbf{X}_r[k]$ is a sparse vector.
\be
\begin{split}
&\widetilde{\mathbf{H}}_r[k]= [\mathbf{h}_r^1[k], \mathbf{h}_r^2[k], \ldots, \mathbf{h}_r^{\mathrm{RF}}[k]], k=1, \ldots, N,\\
&\mathbf{X}_r[k]=[\mathbf{x}_r^1[k], \mathbf{x}_r^2[k], \ldots, \mathbf{x}_r^{\mathrm{RF}}[k]],  k=1, \ldots, N.
\label{HE}
 \end{split}
 \ee
We can represent the effective channel matrix $\widetilde{\mathbf{H}}_r[k]$ as
\be
\widetilde{\mathbf{H}}_r[k]=\mathbf{\Psi}_r \mathbf{X}_r[k], k=1, 2, \ldots, N.
\label{HX}
\ee
Then, $M_r$ training transmissions are needed to obtain the measurement signal at the $k$th subcarrier,
which can be expressed as
\be
\begin{split}
\mathbf{R}_r[k]=&\mathbf{\Phi}_r ( \mathbf{Y}_r[k] + \mathbf{N}_r[k] )\\
=&\mathbf{\Phi}_r (\mathbf{\Psi}_r  \mathbf{X}_r[k]+ \mathbf{N}_r[k]) \\
=&\mathbf{V}_r  \mathbf{X}_r[k]+  \mathbf{\Psi}_r\mathbf{N}_r[k], k=1,\ldots, N,
\label{rr}
\end{split}
\ee
where $\mathbf{R}_r[k] \in \mathbb{C}^ {M_r \times N_{\mathrm{RF}}}$ is the receive signal matrix,
$M_r$ is the length of training signal,
$\mathbf{\Phi}_r \in \mathbb{C}^ {M_r \times N_r}$ is the measurement matrix which is randomly
chosen from the set $\{\pm1,\pm j \} $,
 $\mathbf{N}_r[k]\in \mathbb{C}^{N_r \times N_{\mathrm{RF}}}$ is the AWGN matrix at the $k$th subcarrier,
 $\mathbf{V}_r \triangleq \mathbf{\Phi}_r \mathbf{\Psi}_r $.
During the $m$th training transmission,  $m=1, \ldots, M_r$,
the transmitter use $\mathbf{f}_{\mathrm{RF}_i}$ as the transmit beam vector,
while the receiver uses $\mathbf{\Phi}_r(m,:)^H$ as the beam vector.
To estimate the effective channel $\widetilde{\mathbf{H}}_r[k]$, we can use orthogonal matching pursuit (OMP)
to estimate $\mathbf{X}_r[k]$ and then
the effective channel  $\widetilde{\mathbf{H}}_r[k]$ can be constructed  by (\ref{HX}).

With the estimated effective channel $\widetilde{\mathbf{H}}_r[k]$,
we propose to firstly  design the analog combiner which can enhance the channel gain
of each data stream channel as well as
suppress the interference from each other.
Since each subcarrier has its own optimal precoder/combiner,
it is difficult to choose the best beam vector to achieve the original goal.
Therefore, we turn to seek a suboptimal solution.
In particular, by considering each transmit/receiver RF chain pair one by one,
we successively select analog precoder and combiner to maximize
the corresponding channel gain.

We first calculate the optimal MMSE combiner  to maximize
the corresponding channel gain as well as mitigating the subcarrier interference.
The MMSE combiner for the $k$th subcarrier can be written as
\be
\mathbf{\Gamma}_r[k]=(\widetilde{\mathbf{H}}_r[k]\widetilde{\mathbf{H}}_r^H[k] + \sigma_n^2 \mathbf{I}_{Nr}) ^{-1} \widetilde{\mathbf{H}}_r[k],
\label{MMSE}
\ee
where $\mathbf{\Gamma}_r[k] \in \mathbb{C}^{N_r \times N_{\mathrm{RF}}}, k=1,2,\ldots, N $.
We normalize the MMSE combiner $\mathbf{\Gamma}_r[k]$ by
\be
\big[\mathbf{\Gamma}_r[k] \big] _{:, i} = \frac{\big[\mathbf{\Gamma}_r[k] \big] _{:, i}}{ \left\| \big[\mathbf{\Gamma}_r[k] \big] _{:, i} \right\| }, i=1, \ldots, N_{\mathrm{RF}}.
\ee

For the first data stream channel (i.e. $i=1$),
we  find the suboptimal analog combiner $\mathbf{w}^\star _\mathrm{{RF_1}}$
by searching all candidate vectors in
codebook $\mathcal{W}$ to obtain the largest beamforming gain for all subcarriers:
\be
\mathbf{w}_\mathrm{{RF}_1}^\star = \mathrm{arg} \max_{\mathbf{w}_\mathrm{{RF}} \in \mathcal{W}} \sum_{k=1}^N \big| \mathbf{w}^H_\mathrm{{RF}} \big[\mathbf{\Gamma}_r[k]\big]_{:,1} \big|.
\label{WRF}
\ee
Assign $\mathbf{w}_\mathrm{{RF}_1}^\star$ to the combiner matrices
\be
\big[ \mathbf{W}_\mathrm{{RF}}^\star\big]_{:,1} =\mathbf{w}_\mathrm{{RF}_1}^\star.
\ee

For the rest $N_{\mathrm{RF}}-1$ data streams, we attempt to successively select
combiners to actively avoid the interference of the
data streams whose precoders and combiners have been determined.
To achieve this goal, the component of previous determined
combiners should be removed from other data streams' channels in
such a way that the similar analog  combiners would
not be selected for the other RF chains.
Particularly, let $\mathbf{q}_1 \triangleq \mathbf{w}_\mathrm{{RF}_1}^\star$
be the components of the determined analog
 combinder for the first data stream.
Before finding the second (i.e. $i = 2$) analog combiner,
the MMSE  combiner will be updated  by
\be
\big[ \mathbf{\Gamma}_r[k] \big] _{:, 2} =(\mathbf{I}_{N_r}-\mathbf{q}_1 \mathbf{q}_1^H) \big[ \mathbf{\Gamma}_r[k] \big] _{:, 2},
\ee
and then execute searching precessing as
\be
\mathbf{w}_\mathrm{{RF}_2}^\star = \mathrm{arg} \max_{\mathbf{w}_\mathrm{{RF}} \in \mathcal{W}} \sum_{k=1}^N \big| \mathbf{w}^H_\mathrm{{RF}} \big[\mathbf{\Gamma}_r[k]\big]_{:,2} \big|.
\label{WRF2}
\ee

The analog combiners for the rest RF chains can
be successively selected using the above procedure.
Note that when $i > 1$, the orthogonal component $\mathbf{q}_i$
of the selected combiner  $\mathbf{w}_\mathrm{{RF}_i}^\star$ can be obtained by a
Gram-Schmidt based procedure:
\be
\mathbf{q}_i=\mathbf{w}_{\mathrm{RF}_i}^\star -\sum_{j=1}^{i-1} \mathbf{q}_j^H  \mathbf{w}_{\mathrm{RF}_i}^\star  \mathbf{q}_j,
\ee
\be
 \mathbf{q}_i=\mathbf{q}_i/\| \mathbf{q}_i\|, i=2, \ldots, N_{\mathrm{RF}}.
\ee

%

This  iterative analog combiner design algorithm is summarized in Algorithm 1.

\label{sec:non-iterative}
\begin{center}
\begin{table}[!t]  \hspace{0.5 cm}\small{\textbf{Algorithm 1:} Iterative Analog  Combiner Design}  \vspace{-0.0cm}
\begin{center} \begin{small}
\begin{tabular}{l}
\hline \hline \vspace{-0.2 cm}\\
\hspace{-0.2 cm} \textbf{Input:} \hspace{0.2cm}  $\mathcal{W}$, $\mathbf{\Gamma}_r[k]$,  $\widetilde{\mathbf{H}}_r[k]$, $k=1,\ldots,N$.\\
\hspace{-0.2 cm} \textbf{Output:} \hspace{0.2cm} $\mathbf{W}_\mathrm{{RF}}$.\\  
\hspace{-0.2 cm} \textbf{for} $i=1:N_\mathrm{{RF}}$ \\
\hspace{0.4 cm} $\mathbf{w}_{\mathrm{RF}}^\star (:,i)= \textrm{arg} \underset{\substack{\mathbf{W}_{\mathrm{{RF}}} \in \mathcal{W}}} {\textrm{max}}
\sum_{k=1}^N \big| \mathbf{w}^H_{\mathrm{{RF}}} \big[ \mathbf{\Gamma}_r[k] \big] _{:, i}  \big|$; \\
\hspace{0.4 cm} $\big[ \mathbf{W}_\mathrm{{RF}}^\star\big]_{:,i} =\mathbf{w}_{\mathrm{RF}_i}^\star;$\\
\hspace{0.4 cm} \textbf{if} $i=1$, \\
\hspace{0.4 cm} $ \mathbf{q}_i= \mathbf{w}_{\mathrm{RF}_i}^\star$;\\
\hspace{0.4 cm} \textbf{else} \\
\hspace{0.4 cm}  $\mathbf{q}_i=\mathbf{w}_{\mathrm{RF}_i}^\star -\sum_{j=1}^{i-1} \mathbf{q}_j^H  \mathbf{w}_{\mathrm{RF}_i}^\star
 \mathbf{q}_j=\mathbf{q}_i/\parallel \mathbf{q}_i\parallel$;\\
 \hspace{0.4 cm} \textbf{end if} \\
\hspace{0.4 cm} $\big[ \mathbf{\Gamma}_r[k] \big] _{:, i} =(\mathbf{I}_{N_r}-\mathbf{q}_i \mathbf{q}_i^H) \big[ \mathbf{\Gamma}_r[k] \big] _{:, i}$; \\
\hspace{-0.2 cm} \textbf{end for}\\
\hline
\vspace{-0.99 cm}
\end{tabular}
\end{small}
\end{center}
\end{table}
\end{center}

\vspace{-0.5 cm}
\subsection{Analog Precoder Design with Reverse  Channel Estimation}
With the obtained analog combiner $\mathbf{W}^\star_\mathrm{RF} $ at the receiver,
the channel estimation and hybrid precoder design in reverse channel are similar as
those in forward channel.
The achievable spectral efficiency of the reverse channel system is
\be
\begin{split}
&\mathcal{I}_t =\frac{1}{N}\sum_{k=1}^{N} \mathrm{log}_2 \Big( \big| \mathbf{I}_{N_s} + \frac{P}{N_s} \mathbf{R}_t^{-1} \mathbf{F}_{\mathrm{BB}}^H[k] \mathbf{F}^H_{\mathrm{RF}} \mathbf{H}_r^H[k]  \mathbf{W}_\mathrm{RF}\\
& \times \mathbf{W}_\mathrm{{BB}}[k] \mathbf{W}_\mathrm{{BB}}^H[k]  \mathbf{W}_\mathrm{{RF}}^H \mathbf{H}_r[k] \mathbf{F}_{\mathrm{RF}} \mathbf{F}_{\mathrm{BB}}[k]\big| \Big),  k=1,\ldots, N,
\label{sumrate1}
\end{split}
\ee
where the subscript ``$t$"  represents the transmitter,
 $\mathbf{R}_t \triangleq \sigma_t^2 \mathbf{F}_\mathrm{{BB}}^H[k] \mathbf{F}^H_\mathrm{{RF}} \mathbf{F}_\mathrm{{BB}}[k] \mathbf{F}_\mathrm{{RF}} $
is the noise convariance of the reverse channel.


Since the analog  combiner  $\mathbf{W}_\mathrm{RF} $ is available, the receiver also transmits $M_t$ training symbols to
estimate the  effective  reverse channel.
The received signal at the transmitter is
\be
\begin{split}
\mathbf{Y}_t^i[k]=  \mathbf{H}_r^H[k] \mathbf{w}_{\mathrm{RF}_i} \mathbf{s},  i=1, \ldots, N_{\mathrm{RF}},
\label{yt}
\end{split}
\ee
where $\mathbf{Y}_t^i[k] \in \mathbb{C}^{N_t \times 1}$ denotes the received signal for the $i$th RF chain.
Similar to (\ref{HE}), let $\widetilde{\mathbf{H}}_t[k] $ represent the effective reverse channel and
 we have
 \be \widetilde{\mathbf{H}}_t[k]=\mathbf{\Psi}_t \mathbf{X}_t[k], k=1, \ldots, N, \ee
where $\mathbf{\Psi}_t$ is the dictionary of size $N_t \times L_t$ with $L_t=L_r$,
and it is the same as $\mathbf{\Psi}_r$ except that the term $N_r$ in (\ref{PHIr}) is replaced with $N_t$.
The signal after measuring can be expressed as follows
\be
\begin{split}
\mathbf{R}_t[k]=&\mathbf{\Phi}_t ( \mathbf{Y}_t[k] + \mathbf{N}_t[k] )\\
=&\mathbf{\Phi}_t (\mathbf{\Psi}_t  \mathbf{X}_t[k]+ \mathbf{N}_t[k]) \\
=&\mathbf{V}_t  \mathbf{X}_t[k]+ \mathbf{\Psi}_t\mathbf{N}_t[k],  k=1,\ldots, N,
\label{rt}
\end{split}
\ee
where $\mathbf{R}_t[k]$ is the receive signal  matrix of size $ {M_t \times N_{\mathrm{RF}}}$,
$\mathbf{\Phi}_t \in \mathbb{C}^ {M_t \times N_t}$ is the measurement matrix, and the other notations are similar
to those in (\ref{rr}).
We can obtain the effective channel $\widetilde{\mathbf{H}}_t[k]$ based on CS technique
in a similar way to obtain $\widetilde{\mathbf{H}}_r[k]$.

The MMSE of reverse channel can be written as
\be
\mathbf{\Gamma}_t[k]=(\widetilde{\mathbf{H}}_t[k]\widetilde{\mathbf{H}}_t^H[k] + \mathbf{I}_{Nt}) ^{-1} \widetilde{\mathbf{H}}_t[k],
\label{MMSE2}
\ee
where $\mathbf{\Gamma}_t[k]$ is MMSE matrix of size  $N_t \times N_{\mathrm{RF}}$.
Similar to the determination of the analog combiner $\mathbf{W}^\star_\mathrm{{RF}}$,
the analog precoder $\mathbf{F}_\mathrm{{RF}}$ is selected
by searching through the columns of  codebook $\mathcal{F}$:
\be
\mathbf{f}_{\mathrm{RF}_i}^\star= \mathrm{arg} \max_{\mathbf{f}_\mathrm{{RF}} \in \mathcal{F}} \sum_{k=1}^N \big| \mathbf{f}^H_\mathrm{{RF}}\big[ \mathbf{\Gamma}_t[k]\big]_{:,i} \big|,
i=1, \dots,  N_{\mathrm{RF}}.
\label{FRF}
\ee
Assign $\mathbf{f}_{\mathrm{RF}_i}^\star$ to the precoder matrices
\be
\big[ \mathbf{F}_\mathrm{{RF}}^\star\big]_{:,i} =\mathbf{f}_{\mathrm{RF}_i}^\star, i=1, \dots,  N_{\mathrm{RF}}.
\ee

The following procedure is similar as Algorithm 1  proposed above.
After obtaining the analog precoder $ \mathbf{F}_\mathrm{{RF}}^\star$,
the first iteration is completed.
We then use the obtained analog precoder above as a start point and iteratively design the
analog  precoder and combiner in forward and reverse transmissions.
During the next iteration, a similar process repeats.
After $M_r$ training transmissions from the transmitter,
the receiver will estimate the effective channel $\widetilde{\mathbf{H}}_r[k]$ and then
renew the combiner  $ \mathbf{W}_\mathrm{{RF}}^\star$ by maximizing the spectral efficiency (similar to (\ref{WRF})).
Next, based on the $M_t$ training transmission from the receiver,
the transmitter estimates the  effective channel $\widetilde{\mathbf{H}}_t[k]$
and then updates the precoder by searching through the codebook (similar to (\ref{FRF})).
The iteration procedure  continues until the convergence is found or the iteration times exceeds a pre-specified number.

\subsection{Digital Precoder and Combiner Design}
After all analog beamformer pairs for RF chains have been determined,
 the baseband digital precoder and combiner are
computed to further mitigate the interference and maximize the spectral efficiency.
We can obtain the effective baseband channel as
\be
\mathbf{H}_e[k]= \mathbf{W}_\mathrm{{RF}}^H \widetilde{\mathbf{H}}_r[k], k=1, \dots, N.
\ee
We perform singular value decomposition (SVD)
\be
\mathbf{H}_e[k]=\mathbf{U}_k \mathbf{\Sigma}_k \mathbf{V}_k^H,  k=1, \dots, N,
\ee
where $\mathbf{U }_k$ is an $N_r \times  N_r$  unitary matrix,
$\mathbf{\Sigma }_k$ is an $N_r \times  N_t$  diagonal matrix of singular values,
and $\mathbf{V}_k$ is an $N_t \times  N_t$  unitary matrix.
Then, the digital precoder and combiner can be obtained by
\be
\begin{split}
\mathbf{F}_\mathrm{{BB}}^\star[k] =\mathbf{V}_k (:, 1:N_{\mathrm{RF}}),\\
\mathbf{W}_\mathrm{{BB}}^\star[k] =\mathbf{U}_k (:, 1:N_{\mathrm{RF}}).
\label{digital}
\end{split}
\ee
Finally, the digital precoder and  combiner can be normalized by
\be
\begin{split}
  \mathbf{F}_\mathrm{{BB}}[k] &=\frac{\sqrt{N_s}\mathbf{F}_{\mathrm{BB}}[k]}  {\|\mathbf{F}_\mathrm{{RF}}\mathbf{F}_{\mathrm{BB}}[k]\|_F}, k=1, \ldots, N, \\
  \mathbf{W}_\mathrm{{BB}}[k]& =\frac{ \sqrt{N_{s}}\mathbf{W}_{\mathrm{BB}}[k]}  {\|\mathbf{W}_\mathrm{{RF}}\mathbf{W}_{\mathrm{BB}}[k]\|_F}, k=1, \ldots, N.
\end{split}
\ee

\vspace{0.2 cm}
\section{Simulation Results}
In this section, we illustrate the simulation results of the
proposed iterative hybrid precoder and combiner design.
We consider a mmWave MIMO-OFDM system where both the transmitter and receiver are
equipped with 32-antenna ULAs and the antenna spacing is $d=\frac{\lambda}{2}$.
The number of channel paths is  set as $L=6$.
The elevation of the AoA/AoD is assumed to be uniformly distributed in $[-\frac{\pi}{2}, \frac{\pi}{2}]$.
The number of RF chains at transmitter and receiver  are $N_{\mathrm{RF}}=4$, so is the number of data streams $N_{s}=4$.
We employ two codebooks consist of a series of array response vectors with 64 uniformly quantized angle resolutions.
For the comparison purpose, we also evaluate the approximate Gram-Schmidt based hybrid precoding algorithm \cite{gram},
which greedily selects the RF beamforming vectors using Gram-Schmidt orthogonalization.
Moreover,  the greedy hybrid precoding/combining algorithm is proposed  with the assumption of  perfect CSI.

\begin{figure}[!t]
   \begin{center}
    \vspace{-0.3cm}
        \includegraphics[width= 3.5 in]{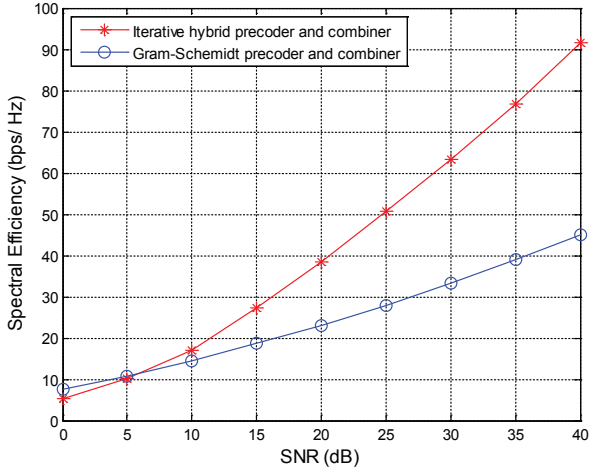}
         \vspace{-0.2 cm}
         \captionsetup{justification=centering}
    \caption{Spectral efficiency versus SNR ($N=32, N_t= N_r=32,  N_{\mathrm{RF}}=4$).}
   \label{P1}
   \end{center}
\end{figure}

\begin{figure}[!t]
   \begin{center}
    \vspace{-0.3cm}
        \includegraphics[width= 3.5 in]{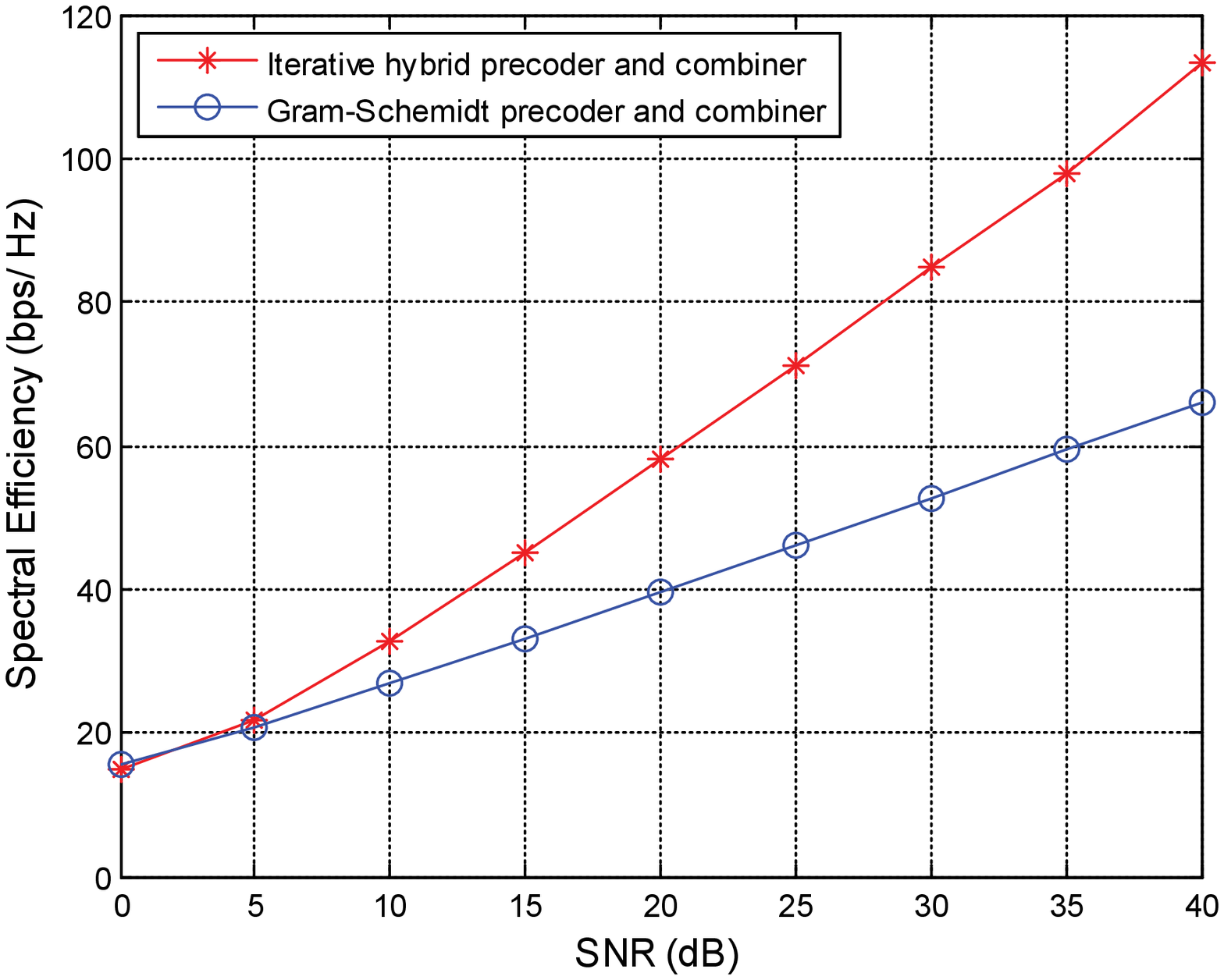}
         \vspace{-0.2 cm}
         \captionsetup{justification=centering}
    \caption{Spectral efficiency versus SNR ($N=128, N_t= N_r=32, N_{\mathrm{RF}}=4$).}
   \label{P2}
   \end{center}
\end{figure}

Fig. \ref{P1}  shows the spectral efficiency versus signal-to-noise-ratio
(SNR) over $10^5$ channel realizations in mmWave MIMO-OFDM system.
The number of subcarriers is $N=32$ and the CP length is $ N_\mathrm{cp}=8$.
We can observe  that the spectral efficiency of our proposed iterative
hybrid precoder and combiner design is higher than the Gram-Schmidt based greedy hybrid precoding and combining algorithm.
A mmWave MIMO-OFDM system with $N=128$ subcarriers and $ N_\mathrm{cp}=32$ is shown in Fig. \ref{P2} which has similar conclusions.
The spectral efficiency of the two algorithms will be improved  with the number of subcarriers increasing.
It can be observed from these two figures that our  proposed algorithm
has better performance compared with the existing algorithm.

\begin{figure}[!t]
   \begin{center}
    \vspace{-0.3cm}
        \includegraphics[width= 3.5 in]{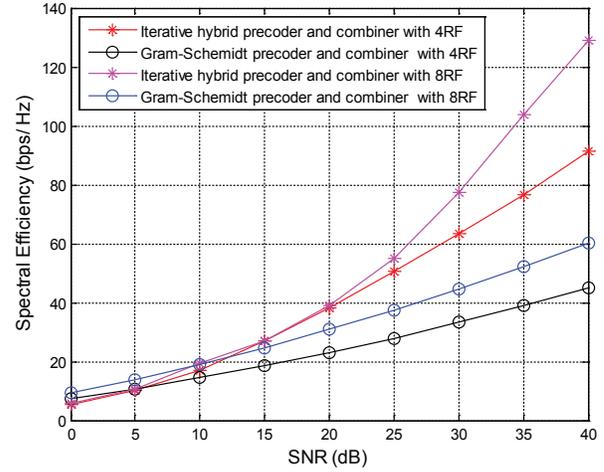}
         \vspace{-0.2 cm}
         \captionsetup{justification=centering}
    \caption{Spectral efficiency versus SNR ($N=32, N_t=N_r=32$).}
   \label{P3}
   \end{center}
\end{figure}

Fig. \ref{P3}  shows the  spectral efficiency versus SNR using different number of RF chains in  mmWave MIMO-OFDM systems.
Similar conclusions can be drawn that our proposed iterative  hybrid  precoding/combining  algorithm
can achieve significant performance improvement.
In addition, our proposed hybrid precoding/combining  algorithm with $N_\mathrm{RF}=8$ has better performance than
the case of $N_\mathrm{RF}=4$.  We can observe that the larger number of RF chains, the better
performance can be achieved in mmWave systems.


%
%
%

\section{Conclusions}
This paper considered the problem of iterative hybrid precoder and combiner design in mmWave MIMO-OFDM systems.
We proposed an iterative scheme to
successively select the analog precoder and combiner in forward and reverse channels.
A CS-based channel estimation is firstly utilized to estimate the effective channel.
Then, based on the effective channel, the analog precoder or combiner can be obtained by OMP algorithm.
Finally, the digital precoder and combiner are computed to
further suppress the interference and maximize the spectral efficiency.
Simulation results demonstrate that our proposed scheme can achieve significant performance improvement.

\end{document}